\documentclass[12pt]{spieman}  % 12pt font required by SPIE;
\usepackage{amsmath,amsfonts,amssymb}
\usepackage{graphicx}
\usepackage{setspace}
\usepackage{tocloft}
\usepackage[english]{babel}
\usepackage[utf8]{inputenc}
\usepackage[T1]{fontenc}
\usepackage{hyperref}

\usepackage[left=2cm,right=2cm,top=2cm]{geometry}
\usepackage{caption}
\usepackage{subfig}
\usepackage{tabularx}
\usepackage{multirow}
\usepackage[toc,page]{appendix}
\usepackage[normalem]{ulem}
\usepackage{pgfplots}
\usepackage{tikz}

\title{BOCOSUR: An all sky network for fireball detection in Uruguay}

\author[a,*]{M. Caldas}
\author[a]{A. Guaimare}
\author[a]{V. Abraham}
\author[a]{L. Barrios}
\author[a]{M. Hern\'andez}
\author[a]{L. Velasco}
\author[a]{G. Tancredi}
\affil[a]{Universidad de la Rep\'ublica, Facultad de Ciencias, Instituto de F\'isica, Departamento de Astronom\'ia, Igu\'a 4225, CP 11400, Montevideo, Uruguay.}

\cftpagenumbersoff{figure}
\cftpagenumbersoff{table} 

\begin{document} 
\maketitle

\begin{abstract}
Over the past couple of decades, several networks for the automatic detection of fireballs have been deployed. Their primary scientific goal is to facilitate the rapid recovery of meteorites, determine their pre-atmospheric orbits, and look for possible dynamic links with parent bodies. The Bocosur network is a contribution to the global deployment of automated fireball networks and to the increase of the number of recoverable meteorite falls. It is located in Uruguay, South America (Lat: -30$^{\circ}$ to -35$^{\circ}$). Its main scientific goal is the detection of fireballs of asteroidal origin, massive enough to produce meteorites, and also to inspire secondary-level students and teachers through their involvement in this citizen-science oriented project. The deployment of this network started in 2019, and was completed in March, 2023, when we installed 20 stations separated $\sim 120$ km, covering an area of $\sim 180,000$ km$^2$. During this period of time, one major technological upgrade was made when we migrated from a well-known camera to a higher-resolution, more sensitive system. We were able to build a completely autonomous system at an affordable cost that can be replicated in all the stations. A comparison between the astrometric and photometric performance of these two detection systems is reported. Also, a photometric methodology for estimating the brightness of very bright fireballs is presented and validated against the known magnitudes of Jupiter and the full Moon. We obtain mean residuals of the astrometric reduction of $\sim$5', and the discrepancy between the obtained brightness of Jupiter and the Moon average to 0.18 and 1.2 magnitudes, respectively. Results on the processing of a very bright (M$_{peak}\sim$-9.0 mag) fireball detected in four stations are also presented.  
\end{abstract}

% Include a list of up to six keywords after the abstract

% Include email contact information for corresponding author
{\noindent \footnotesize\textbf{*}Manuel Caldas,  \linkable{manuel.caldas@fcien.edu.uy} }

\begin{spacing}{2}   % use double spacing for rest of manuscript

\section{Introduction}\label{sec:intro}
Meteorites constitute a valuable source of information on the formation and evolution of the Solar System, including the processes that took place during the early formation of planetesimals and planets. A rapid recovery of fallen material is essential in retrieving a fresh and unweathered sample, which, together with the determination of the meteoroid's pre-atmospheric orbit, can significantly constrain the conditions on its origin. 

Meteoroids that lead to meteorite falls come mainly from Main Belt Asteroids (MBAs) and from parent objects in Earth crossing orbits \cite{Wetherill1981}. Although the number of meteorites with computed orbits is fairly low (47 as of December, 2023, which includes the most recent Antonin, Winchcombe, Traspena and \r{A}dalen meteorites, plus 43 mentioned in \cite{Andrade_2023}), some pattern appear to emerge from this data.

As pointed out in \cite{Jenniskens_2020}, LL chondrites appear to originate in the inner MAB ($\nu_6$ resonance); CM chondrites appear to be linked to a source which efficiently pumps material into the 3:1 mean-motion resonance with Jupiter; H and L chondrites originate from diverse sources, linked to the 3:1 or the $\nu_6$ resonances, as indicated by their Cosmic-Ray Exposure (CRE) ages in combination with their dynamical history (see \cite{Jenniskens_2020} and references therein).

\renewcommand{\thefootnote}{\roman{footnote}}

This tiny fraction of meteorites with computed orbits over the total number of available meteorites (more than 70.000, as registered in the Meteoritical Bulletin Database \cite{MetBullDB}) , is eloquent in the need of increasing its number for a better understanding of their nature, origin and classification. 

The most efficient and cost-effective way of accomplishing this is by deploying networks of automated detection systems, with a node separation that enables a simultaneous detection in at least 3 stations. \cite{Devillepoix_2020} estimate that for an accurate calculation of a meteorite strewn field to ensure successful recovery, at least one station should be no more than 130 km away from the fall site, and a second station should be no more than 300 km away. Considering that most of fireball phenomena occur at heights between of 50 and 100 km, and each all-sky camera covers a full angle of 140$^{\circ}$ (the astrometric and photometric quality deteriorates significantly at altitudes below 20 $^{\circ}$, due to distortions close to the lens' boundaries, and high air-masses, respectively), usual average separation between stations is $\sim$100-150 km (see e.g. \cite{Colas2020}$^,$ \cite{Pena-Asensio_2021}$^,$\cite{Howie_2017}).   

Several such networks exist today and most of them were deployed in the last decade and a half: FRIPON \cite{Colas2020}, PRISMA \cite{Gardiol2016}, Southern Ontario \cite{Weryk2008}, the Desert Fireball Network (DFN) \cite{Devillepoix_2020}, latest upgrades to the European Fireball Network \cite{Spurny2017}, the Spanish Meteor Network (SPMN) \cite{Trigo-Rodriguez_2004} and the Brazilian Meteor Observation Network \cite{Amaral_2018}. Also, as of August, 2023, seven stations belonging to the FRIPON network have been deployed in Chile, and some more are pending to be installed \cite{FriponCL}. 
The biased geographical distribution of these networks is eloquently reflected in the fact that only 8 of the 46 meteorites with computed orbits were found in the Southern Hemisphere, 5 of which were recovered in Australia, and only two were found in Latin America (Cuba and Brazil). This justifies the need for deploying similar infrastructure in the Southern Hemisphere, and particularly in South America. 

The growing number of networks and the development of existing ones demands the automation of as much stages as possible in the data processing pipelines. There are several software options for automation of the first stages that involve the detection itself and some data pre-processing. 

As indicated by \cite{Weryk_2013}, examples of these are MeteorScan \cite{Gural_1997}, MetRec \cite{Molau_1999}, ASGARD \cite{Weryk2008} and UFOCapture. More recently, the FreeTure package developed by the FRIPON team is a freely available option, and is capable of controlling both GigE cameras and some USB 2.0 and 3.0 cameras \cite{Colas2020}. Recent advancements in automated meteor detection tools have enabled some of networks to operate with a high degree of autonomy, from acquiring data to reliably detecting meteors (see e.g. \cite{PenaAsensio2023} and references therein).

Less extensive is the list of automated options for the analysis of detected meteors. For instance, the DFN uses an implementation of the so called Dynamic Trajectory Fit (DTF) method described in \cite{Jansen-Sturgeon_2020}, publicly available through the network’s github page. Also, software not specifically developed for all sky meteor data is being used, such as SExtractor for pixel identification of stars, and Scamp for a first-order astrometric reduction, suitable for elevations above $\sim$30$^{\circ}$, which then feeds the initial conditions of more suited models for all sky data \cite{Jeanne_2019}. A package that tackles all stages from detection to 3D trajectory and orbit reconstruction is FireTOC, developed by the SPMN team \cite{Pena-Asensio_2021}.

In this paper, we present the BOCOSUR network, a set of locally developed automated all-sky detection systems deployed in Uruguay (South America), for the detection and post-processing of fireball events. With a dense network of all-sky cameras, we are able to get a multi-station detection of fireballs, which allows us to obtain a precise pre-atmospheric trajectory of the meteoroid, and calculate the strewn-field information for achieving the recovery of new meteorite fall.

In Section \ref{sec:data}, we describe the overall network and the main features of its hardware and data processing pipeline. Section \ref{sec:methods} describes the algorithms and methodology employed in the pre- and post-processing of events, particularly the astrometric and photometric reduction procedures. Section \ref{sec:results} presents the main results on astrometric and photometric performance of our system and proposed methodologies, and the processing of a fireball detected in five stations, as well as the outlining of future perspectives. Conclusions are contained in Section \ref{sec:concl}.

\section{Network description}\label{sec:data}
The BOCOSUR project is an initiative driven by the Department of Astronomy of the Faculty of Sciences of the University of the Republic of Uruguay (Udelar)\cite{BocosurW}.  After many years developing different prototypes of all-sky cameras, the project formally started in 2019; its goal is to deploy a nationwide network of all sky systems for automatic detection of fireballs, focused on very bright fireballs of asteroidal origin that may give rise to meteorites. 

The network deployment was completed in March, 2023, and all 20 stations comprise an acquisition system based on a ZWO ASI 178MM camera, which has a 6.4 Mpix CMOS sensor. Initially, the broadly used Watec 902 H2 Ultimate CCTV camera was used, but was successively substituted in the different stations during the first semester of 2022. The main reason for this upgrade is the improved sensitivity and resolution, as described in Section \ref{sec:comp}.

\begin{figure}[!ht]
\centering
\includegraphics[width=\textwidth]{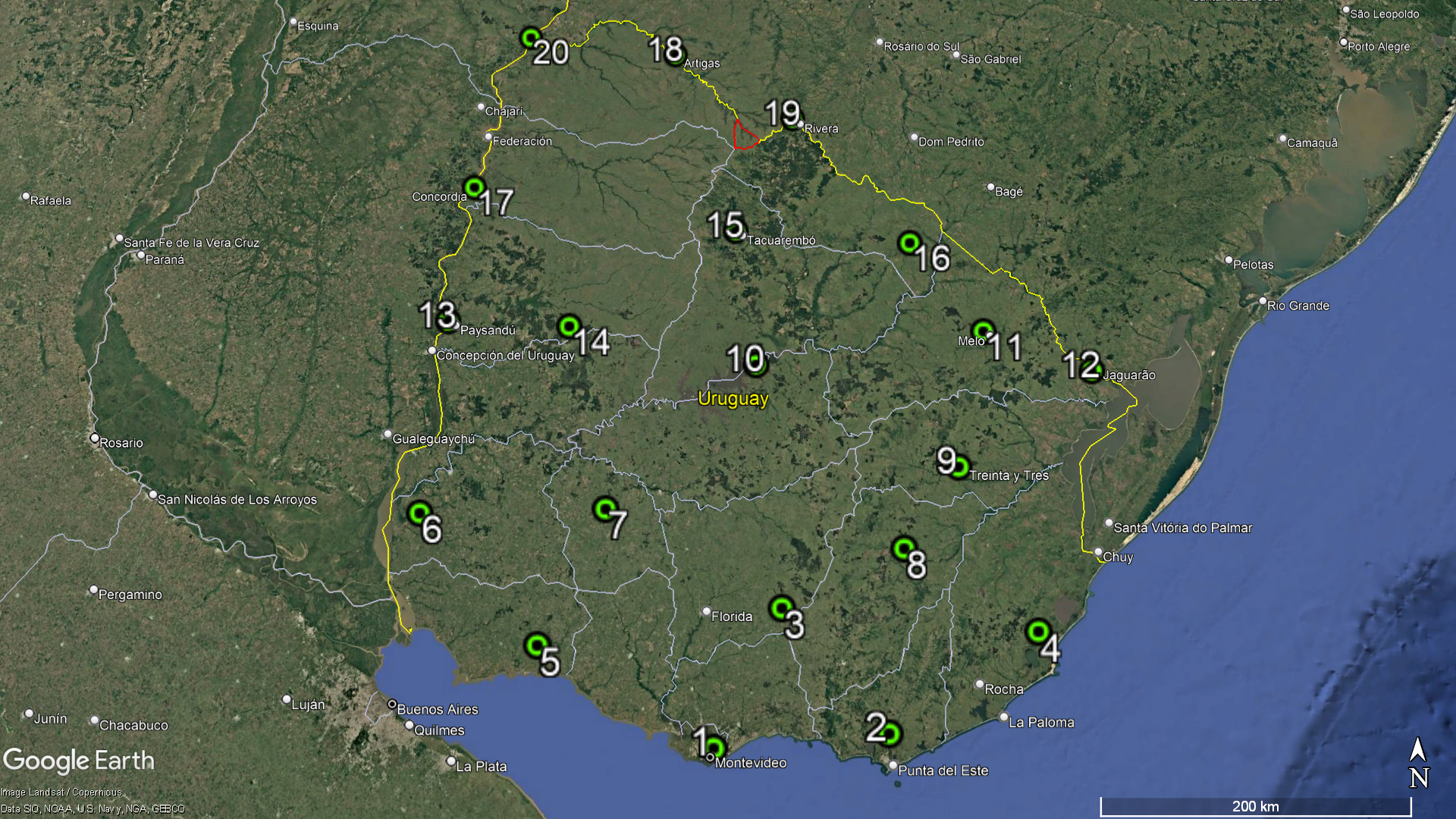}
\caption{Location of the 20 stations of the BOCOSUR network, in Uruguay (source of the map is Google Earth Pro$\copyright$ v7.3.6.10201, based on Landsat/Copernicus imagery of 12/31/2020, eye alt. 1047.88km. SIO, NOAA, U.S. Navy, NGA, GEBCO).}
\label{fig:mapa}
\end{figure}

Figure \ref{fig:mapa} shows the location of the stations and Table \ref{table:estaciones} lists the educational institutes where they are located, as well as their geographic coordinates. Most of the stations are installed on the roofs of secondary schools, and the reason for this is that the project is intended to have a strong citizen science profile, involving teachers and students in the operation of each station. Also, several workshops have been held by the project's staff to instruct the participants (again, both secondary school teachers and students) in the manual classification of videos. This classification has been used, in turn, to train an automatic video classifier, as will be reported in an ongoing paper. 

The following sections describe the hardware and software used for detection and data processing. 

\begin{table*}[!ht]
\small
\centering
\caption{List of educational institutions where the stations are installed (\emph{liceo} is the Spanish word for secondary school). Latitude and Longitude are expressed in degrees and minutes with decimals.}
\label{table:estaciones}
\begin{tabular}{|c|l|l|c|}
\hline
\emph{ID} & \emph{Institute} & \emph{City, Department} & \emph{(Lat,Lon)}\\
\hline

1   & Facultad de Ciencias, Udelar  & Montevideo & (-34 52.92, -56 \ 7.05) \\
2   & Liceo 1 “Monseñor Mariano Soler”   & San Carlos, Maldonado & (-34 47.49, -54 55.05) \\
3   & Liceo “Ram\'on Goday”   & Casup\'a, Florida & (-34 \ 5.92, -55 39.00) \\
4   & Liceo “Jos\'e Aldunate Ferreira”  & Castillos, Rocha & (-34 12.02, -53 51.73) \\
5   & Liceo “Agust\'in Urbano Indart Curuchet”  & Rosario, Colonia & (-34 18.74, -57 20.72) \\
6   & Liceo 1 “Dr. Roberto Taruselli”  & Dolores, Soriano & (-33 31.85, -58 13.01) \\
7   & Liceo 1 “Carlos Brignoni Mosquera”   & Trinidad, Flores  & (-33 3.122,  -56 53.75) \\
8   & Estaci\'on Agraria, UTU  & Piraraj\'a, Lavalleja  & (-33 44.39, -54 47.29) \\
9   & Liceo 1 “Dr. Nilo L. Goyoaga”  & Treinta y Tres, Treinta y Tres & (-33 14.31, -54 23.12) \\
10  & Liceo “Lucio Gabino N\'uñez”  & San Gregorio de Polanco, Tacuaremb\'o & (-33 31.85,  -58 13.01) \\
11  & Liceo 1 “Juana de Ibarbourou”  & Melo, Cerro Largo & (-32 21.99, -54 10.19) \\
12  & Liceo “Dr. An\'ibal Acosta Estap\'e”   & R\'io Branco, Cerro Largo & (-32 35.89, -53 23.06) \\
13  & Liceo 1 “Qu\'imica Farmac\'eutica Elida Heinzen”  & Paysand\'u, Paysand\'u & (-32 18.93, -58 \ 5.59) \\
14  & Liceo de Guich\'on  & Guich\'on, Paysand\'u & (-32 2.184, -57 11.99) \\
15  & Liceo 1 “Ildefonso Pablo Est\'eves”  & Tacuaremb\'o, Tacuaremb\'o & (-31 42.88, -55 59.20) \\
16  & Liceo de Vichadero  & Vichadero, Rivera  & (-31 46.70, -54 41.47) \\
17  & Liceo 1 “Instituto Polit\'ecnico Osimani y Llerena”  & Salto, Salto &  (-31 23.41, -57 57.46) \\
18  & Liceo 2  Artigas & Artigas, Artigas & (-30 24.32, -56 27.64) \\
19  & Liceo 5  "Prof. Carlos Ma. Thieulent" & Rivera, Rivera & (-30 53.63, -55 33.47) \\
20  & Liceo 1 “Escribano Diego Carlos Muguruza”  & Bella Uni\'on, Artigas & (-30 15.37, -57 35.42) \\
\hline
\end{tabular}
\end{table*}

\subsection{Hardware}
\begin{figure}[htpb]
    \begin{minipage}{\linewidth}
        \centering
        \includegraphics[height=0.3\textheight]{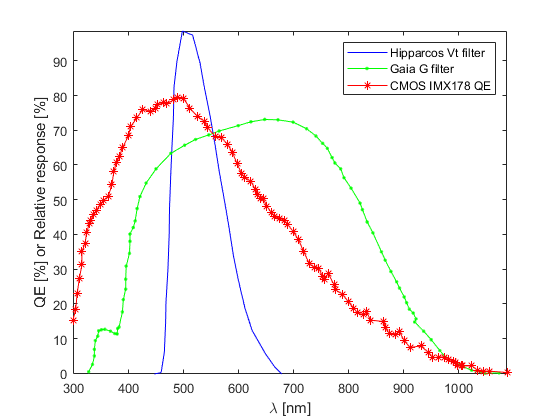}
        \caption[Quantum efficiency curve of the CMOS IMX178 sensor (red curve, digitized from Figure 2 in \cite{Bailey_2023}), transmissivity of the Gaia (E)DR3 G passband (green), and spectral response of Hipparcos' $V_T$ filter (blue, digitized from Figure 1.3.1 in \cite{Perryman_1997b}).]{Quantum efficiency curve of the CMOS IMX178 sensor (red curve, digitized from Figure 2 in \cite{Bailey_2023}), transmissivity of the Gaia (E)DR3 G passband (green\footnote{https://www.cosmos.esa.int/web/gaia/edr3-passbands}), and spectral response of Hipparcos' $V_T$ filter (blue, digitized from Figure 1.3.1 in \cite{Perryman_1997b}).}   
        \label{fig:qe}        
    \end{minipage}
\end{figure}

The acquisition system is based on the ZWO ASI 178MM (Mono) camera, which has a 1/1.8'' CMOS sensor (IMX178) and a 14-bit ADC. The sensor's peak quantum efficiency (QE) is 81\%, as shown in Figure \ref{fig:qe}. The highest available video resolution is 2080x3096 pixels.

% https://www.cosmos.esa.int/web/gaia/edr3-passbands

\begin{figure}[!ht]
\centering
\includegraphics[height=0.4\textheight]{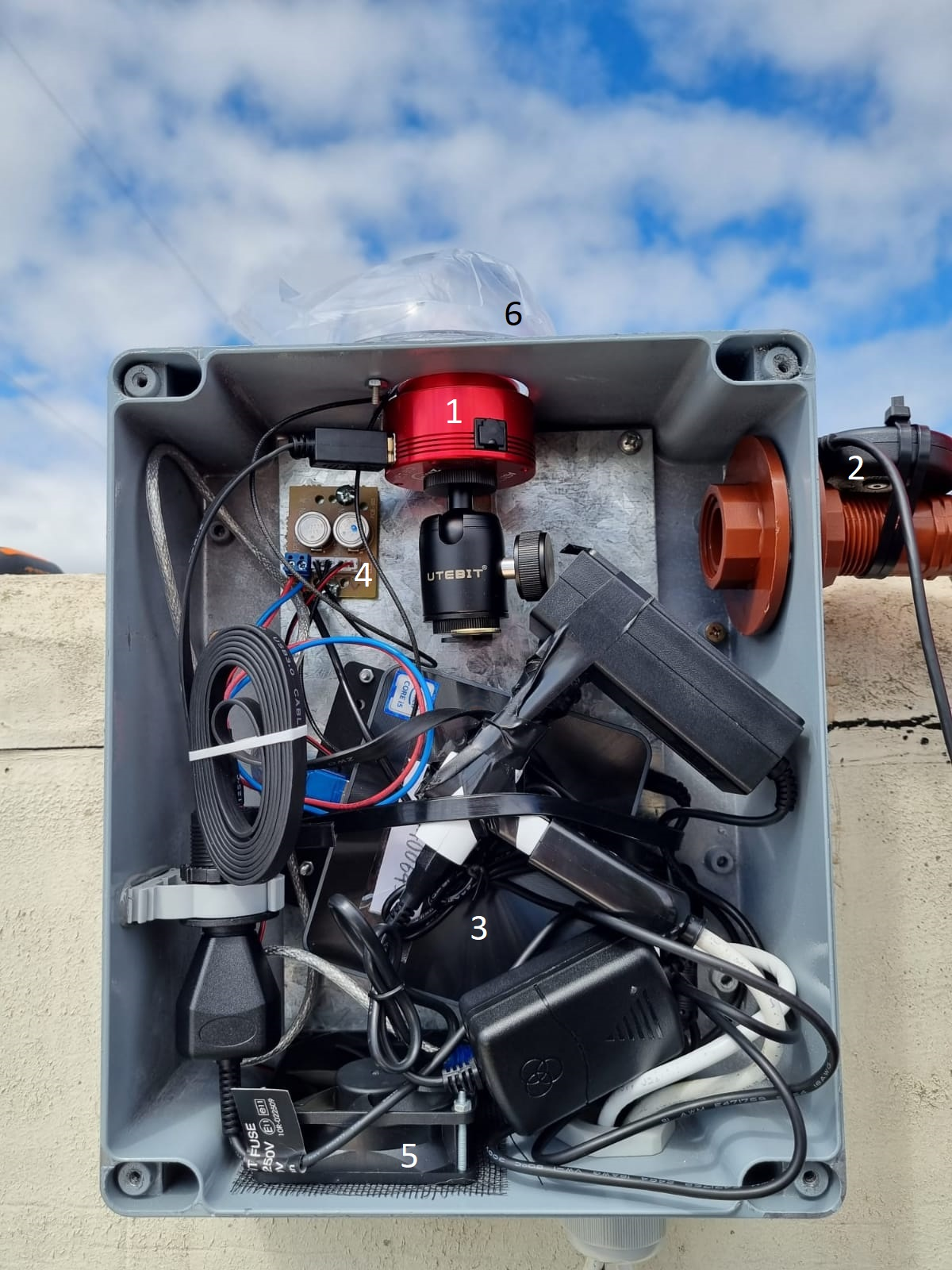}
\caption{Locally developed all sky system, consisting of: (1) an ASI 178MM camera connected to a mini PC (3); a GPS dongle (2), which provides accurate time and position; a CCTV heater and cooler system (4) which controls a fan (5) and a $\sim$3W resistor (not shown), placed on top of the enclosure and under a hard-coated polycarbonate dome (6).}
\label{fig:hw_in}
\end{figure}

The optics consists of a 2.5mm fisheye lens, which provides a $\sim$170$^{\circ}$ field-of-view (FOV), and a hard-coated polycarbonate (3.85'' outside diameter) dome on top of camera and lens. The camera is connected via USB3.0 to a local nano PC (i5-8279U CPU, 2.40GHz, 16 GB RAM) running Windows 10. All this equipment, together with a CCTV heater and cooler (fan) is enclosed within a waterproof (IP 67) aluminum enclosure (see Figure \ref{fig:hw_in}). Additionally, a GPS receiver is connected to the PC to provide accurate time, as explained in \ref{sec:pipeline}. Recent improvements include a temperature, humidity and heater monitoring module based on an Arduino Nano board and a DHT11 sensor. The stations are connected via ethernet to the institutes internet network. The electric plug from the station is connected to a Wi-Fi smart switch, with which we can remotely switch on and off the computer and make a hard reboot when needed.

The assembling of the stations was done in our laboratory by our team. The first station was installed in mid-2019, followed by two more in early 2020. Then, due to the pandemic, we had to defer the installation to late 2021 for 4 more stations. The remaining 13 were installed between October, 2022 and February, 2023. All the installations were carried out by our team, with the support of teachers and technical staff at each institute.

All stations have fiber optic connectivity, which is possible thanks to the high penetration rate of Fiber To The Home (FTTH) in Uruguay. This provides excellent internet connectivity for educational institutes, which are densely distributed geographically. These two factors made it possible for us to find educational institutions $\sim$ 120 km apart, with excellent connectivity, to install the stations for our fireball network.

\subsection{Software package developments}
All of the applications and scripts for camera control, calibration, data acquisition, data transfer and post-processing have been locally developed in MATLAB (applications) and Python (scripts). 

The application for camera control, calibration and data acquisition runs locally on each PC, and enables to set all the camera properties made available by the manufacturer through the camera's DirectX driver (relevant for our purposes are gain and exposure). Its source code is fully available in \cite{github_bolidosGUI} and an installer for Windows 10 is also provided in the repository.

To mask out unwanted objects from the region of interest - ROI (city lights, trees, antennae, buildings, etc), a customized horizon is defined by taking a snapshot and manually delimiting the mask on the image. It is possible to manually record videos of fixed lengths (5, 10, ..., 60 sec), i.e for later astrometric calibration.  

All parameters that control the detection algorithm (explained in Section \ref{sec:algor}) are settable in the main GUI window, as well as the Sun's elevation angle below which the detection algorithm starts running (possible values are -5, -10 and -15  degrees).  

Additionally, an application for post-processing detected events has been developed and is also available in \cite{github_postproB}. With this application, the user can perform both astrometric and photometric calibrations, and apply them to a fixed (e.g. Moon, planets, stars) or moving (meteor) body on the video stream. All astrometric and photometric results reported in this article were obtained using this software, and the methodologies involved are explained in Sections \ref{sec:astrocalib} and \ref{sec:photocalib}.

\subsection{Data processing pipeline}\label{sec:pipeline}
The detection application (\emph{bolidosGUI}) generates three files for every detected event, all in a local folder Events\textbackslash jjjj\textbackslash dd, where jjjjdd is the Julian date at the beginning of the night when the event took place. 

The first file is a MATLAB metadata file (.mat), having a filename of the form \emph{Event\_yyyy-mm-dd-HH-MM-SS}, where the timestamp corresponds to the CPU time of the first acquired frame. This file is a structure containing, among other parameters, the absolute time of each individual frame, which is critical for further processing of the meteor, the difference between GPS and CPU times at the time the event was created, and a flag indicating whether the GPS data reading was successful or not. 

The second file is the video file itself, which is a Motion JPEG avi file with maximum video quality, as specified by the \emph{Quality} instance of MATLAB's \emph{VideoWriter} class, having a filename of the form \emph{Station\_X\_yyyy-mm-dd-HH-MM-SS}, where X is the station's ID (1,2,...,20), and the timestamp is also that of the first acquired frame.

Finally, a new row is added to a spreadsheet (.xlsx) named \emph{Planilla\_jjjjdd} (\emph{Planilla} is the Spanish word for spreadsheet), each time a new event is registered. This spreadsheet is generated for later manual classification by groups of secondary school students and teachers, which mark with an X the corresponding type of event, as shown in Table \ref{table:planilla}. Other classes included, not shown in this table for better readability, are "cloud", "lightning", "rain", "city lights" (due to an insufficiently well established mask), "other" and "nothing". The events corresponding to the latter type are basically generated by noise triggering. 

The overall network topology is: i) the stations run the \emph{bolidosGUI}, and a script to monitor and report the status to the server; ii) the server receives the reports and produce a briefed report for the website; it also produces the night time lapse every morning; iii) The recorded video are transferred from the stations to the server for later inspection
(see the Supplementary Material for more information).

\begin{table*}
\centering
        \caption{Example of spreadsheet generated by \emph{bolidosGUI}, and completed by a group of human classifiers (see text for omitted classes).}
        \label{table:planilla}
        \begin{tabular}{|c|c|c|c|c|c|c|}
        \hline
        Video name & Date of classification & Group name & Meteor & Airplane & Animal & $\cdots$ \\
\hline
Station\_1\_2022-05-21-23-52-12& 2022-06-15 & School 1 & X &  &  & \\
Station\_1\_2022-05-22-02-22-46& 2022-06-15 & School 1 &  & X &  & \\
\hline
        \end{tabular}
\end{table*}

\section{Methods}\label{sec:methods}
Next follows a detailed description of the detection algorithm implemented in the \emph{bolidosGUI} app. The astrometric calibration procedure is described in \ref{sec:astrocalib}. The photometric calibration and how photometry is performed on detected fireballs is described in \ref{sec:photocalib}. 

\subsection{Detection algorithm}\label{sec:algor}
The motion detection algorithm is based on the image difference between two consecutive frames. The time difference between the two images to be compared is settable, and a heuristically found value of 0.4 s is set in all stations. 

On the image difference we compute the number ($nDiff$) of pixels having a brightness above a certain threshold ($thr$), and their mean distance to their centroid ($rm$). Two conditions have to be simultaneously fulfilled to trigger an event:
\begin{itemize}
    \item  $nDiff\geq nThr$, where $nThr$ is the user-defined minimum number of pixels above the threshold $thr$. 
    \item $rm\leq Rm$, where $Rm$ is a user-defined upper threshold on mean distance to centroid. This condition minimizes triggering due to brightness changes covering large areas (i.e. cloud displacements, raindrops). 
\end{itemize}
As long as both conditions are fulfilled, the algorithm continues to acquire frames to the memory buffer, up to a certain limit to avoid memory overflow. This limit is also settable in terms of maximum video duration in seconds (20s by default). 

A suitable value on $Rm$ follows from the fact that the average radius of pixels inside a circle of radius $R$ is $2R/3$. In our images, a full moon is approximately a bright circle of radius $\sim45$ pixels. Since very bright fireballs can be as bright as a full moon, an upper threshold of $Rm=2R/3=30$ pixels should be adequate. In actuality, a somewhat higher value of $Rm=40$ pixels is used as a default value.

Once the detection conditions cease to be valid, the files described in \ref{sec:pipeline} are created. Motion detection goes on as long as the Sun's elevation angle is lower than the user-defined threshold.     

\subsection{Astrometric calibration}\label{sec:astrocalib}
The mapping function used to perform astrometric calibrations is a two-valued function of two variables:
\begin{equation*}
F(x,y)=(z,Az)    
\end{equation*}

that transforms pixel coordinates $(x,y)$ on the image, to horizontal coordinates zenith angle ($z$) and azimuth ($Az$).

Several models are available (e.g. \cite{Ceplecha1987}$^,$\cite{Borovicka1992}$^,$\cite{Bannister2013}), the main difference between them being the relationship between distance from sensor center and zenith angle measured from the optical axis. Here we follow the approach described in \cite{Bannister2013}, which implies following set of equations to go from $(x,y)$ to $(z,a)$:

\begin{equation}\label{eq:erreDef}
    r=\sqrt{(x-x_0)^2+(y-y_0)^2}
\end{equation}
\begin{equation}\label{eq:u}
    u=p_0+p_1\cdot r+p_2\cdot r^2
\end{equation}
\begin{equation}\label{eq:be}
    b=a_0-E+\arctan\frac{y-x_0}{x-y_0}
\end{equation}
\begin{equation}\label{eq:zeta}
    \cos{z}=\cos{u}\cdot \cos{\epsilon}-\sin{u}\cdot \sin{\epsilon}\cdot \cos{b}
\end{equation}
\begin{equation}\label{eq:az}
    \sin{(a-E)}=\sin{b}\cdot \sin{u}/\sin{z}
\end{equation}

In Equations \ref{eq:erreDef}-\ref{eq:az}, the plate constants to be determined through the calibration procedure are: $x_0$ and $y_0$, coordinates of the center of projection (COP); $p_0$-$p_2$, coefficients of the polynomial which relates distance from center and angle from optical axis, which is related to the lens' mapping function; and the angles $a_0$ (rotation of the sensor’s x-axis from cardinal south), E (angle between the x-axis and a vector defined by the COP and the true zenith projection point) and $\epsilon$, the angle between the COP and true zenith \cite{Bannister2013}.

The problem of finding these coefficients is a non-linear least squares problem, where we seek to minimize the errors:

\begin{equation}
    |z_i-z_{i,CAT}| 
\end{equation}
\begin{equation}
    |(a_i-a_{i,CAT})\cdot sin(z_{i,CAT})|
\end{equation}

where ($z_i,a_i$) are horizontal coordinates obtained with Equations \ref{eq:erreDef}-\ref{eq:az} applied to pixel coordinates $(x,y)$ of known stars, with corresponding catalogue values ($z_{i,CAT},a_{i,CAT}$). As explained in \cite{Bannister2013}, an explicit expression for a (Eq. \ref{eq:az}) must consider the value of cos(a-E) to produce a four-quadrant result.

The first step of identifying star pixel coordinates is made in our procedure in a semi-automatic manner, by manually adjusting two parameters controlling a first-order transformation, that produces a synthetic star map from a list of the 300 brightest stars in the Hipparcos catalogue \cite{PERRYMAN97}. Then, star positions on the frame are manually indicated and if reasonably close to the synthetic stars, these are correctly and automatically identified.

This procedure can be performed on different video acquisitions, of different nights even, to generate a list of $(x,y,z,a)$ values, which are then used in a non-linear solver. In our case, we use the Levenberg-Marquardt \cite{More78} algorithm through MATLAB's \emph{lsqnonlin} function.  Finally, an outlier iterative removal process is applied until a user-specified overall accuracy (typically equal to the pixel scale) and/or a minimum number of datapoints ($\sim$150-200) is reached.

\subsection{Photometric reduction}\label{sec:photocalib}
Once an astrometric calibration is available, photometry reduction can be performed by identifying stars present in an integrated frame. This is done automatically in our algorithm, by searching for catalogued stars (again, using the list from Hipparcos) above a certain user-specified altitude (typically 10$^{\circ}$), in the close vicinity of calculated star positions.

This identification generates a list of ($x$,$y$,$m_V$,$z$) datapoints. Next, circular aperture photometry is performed automatically on the ($x$,$y$) positions, with $r_*$ typically equal to 1 or 2 pixels, and {annulus radius $r_{bgd}$ typically between 6 and 10 pixels},  for computing the averaged background sky flux, $\langle \mathcal{F}_{bgd} \rangle$. The inner separation between the annulus and the star aperture is fixed and equal to 1 pixel.  

% The computed flux is obtained through Equation \ref{eq:flujo}.

% \begin{equation}\label{eq:flujo}
%     \mathcal{F}_* = \iint_{ap.} (\mathcal{F}(x,y)- \langle \mathcal{F}_{bgd} \rangle ) \,dx\,dy 
% \end{equation}

We perform a first order photometric reduction to obtain the atmospheric extinction coefficient, $k$, and zero point $m_0$.

% , included in Equation \ref{eq:mV}, where $X$ is the airmass.

%  \begin{equation}\label{eq:mV}
%     m_V = -2.5 \cdot \log _{10}(\mathcal{F}_{*}) -k \cdot X + m_0
% \end{equation} 

The procedure is similar to the principle used in the astrometric reduction described in \ref{sec:astrocalib}.In this case, we seek to minimize the errors
\begin{equation}
   \epsilon=|m_V-m_{V,cat}| 
\end{equation}
in a least squares sense. Here we also use MATLAB's implementation of the Levenberg-Marquardt algorithm in the \emph{lsqnonlin} function. The  iterative outlier removal step is performed until at least typically 20 datapoints are used or 0.01 mag overall accuracy is reached. Additionally, only stars having a S/N above a user-specified threshold (typically 2-3) are considered. Here, we use following definition of S/N:

 \begin{equation}
    S/N = \frac{\mathcal{F}_{*}}{\sigma_{bgd} \cdot \sqrt{n_{pix}}}
\end{equation} 

The main goal of the photometric reduction is to estimate the brightness of bright fireballs. This implies an extrapolation from star magnitudes to  $m_V<-4$. In order to estimate the sky-subtracted flux of a fireball, which is a saturated, extended object in the image, we fit a 2D-Gaussian to the brightness, which extrapolates the flux beyond the saturation level. Then, this 2D-Gaussian is space-integrated and a flux is computed, which is converted to magnitude using the previously obtained photometric reduction. 
If planets are present in the field, they are included in the photometric reduction process.

\section{Results and discussion}\label{sec:results}
In subsection \ref{sec:comp}, we present a comparison between the results of astrometric and photometric reduction applied on two co-localized cameras in Station 1 of the BOCOSUR network (Faculty of Sciences, in Montevideo). One camera is a Watec 902 H2 Ultimate, which as indicated in Section \ref{sec:data} was previously used in the network. It is a widely used analog, low resolution and highly sensitive video camera. The second one is a high resolution  digital cameras (ZWO ASI 178MM), currently in use in all stations. The results obtained here were the main motivation for this upgrade.   

In subsection \ref{sec:valid} we present the results of extrapolating photometric calibrations performed, which, as described previously, is done using stars, to bright objects of known brightness, namely the full Moon and Jupiter, in order to quantify the performance of such an extrapolation to bright fireballs.   

Finally, in subsection \ref{sec:firstdet} we apply our data processing pipeline on a bright fireball detected in five of our stations, on October 29, 2022.

\subsection{Comparison between Watec and ASI cameras}\label{sec:comp}
\emph{Astrometric comparison}. The astrometric reduction procedure described in \ref{sec:astrocalib} was performed using a set of videos acquired between 2022/09/26-2022/10/02, for the Watec camera, and 2022/09/28-2022/09/29 for the ASI camera. In the case of Watec, the highest available resolution was used (576x720, standard PAL format), and in total 17 videos were needed to generate a list of 159 datapoints. 

In the case of ASI, eventhough the video acquisition uses the highest available resolution (2080x3096), the videos of registered events are downsized to 1040x1548 pixels using bicubic interpolation, through MATLAB's built-in function \emph{imresize}. This is considered to be a satisfactory compromise between file size and resolution. In this case, 8 videos were used from which 306 datapoints were extracted for the reduction.

We use a list of the 300 brightest stars from the Hipparcos star catalogue \cite{PERRYMAN97} for our reduction, and performed the reduction for the whole dataset. Next, we set the maximum mean absolute error (MAE) threshold in the iterative reduction process to equal the  average pixel scale (15 and 6 arcmin for Watec and ASI, respectively). In other words, the subset of datapoints used in this second reduction, yields MAE of about one pixel. 

\begin{figure}[!ht]
\centering{\includegraphics[width=\textwidth]{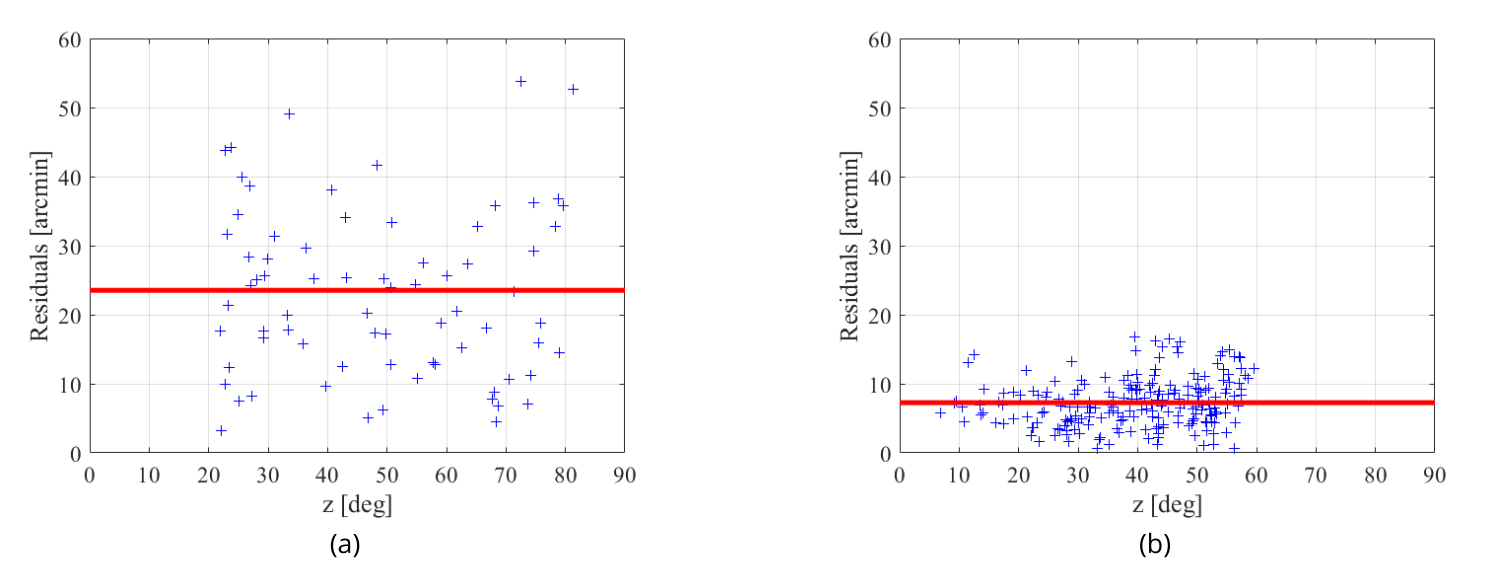}}
\caption{Residuals of the astrometric reduction (as given by Eq. \ref{eq:res_astrom}) for Watec (a) and ASI (b) cameras. The red line indicates the median of the residuals, and is equal to 22.3 and 6.9 arcmin, respectively.}
\label{fig:residuos_astrom}
\end{figure}

\begin{figure}[!ht]
\centering{\includegraphics[width=\textwidth]{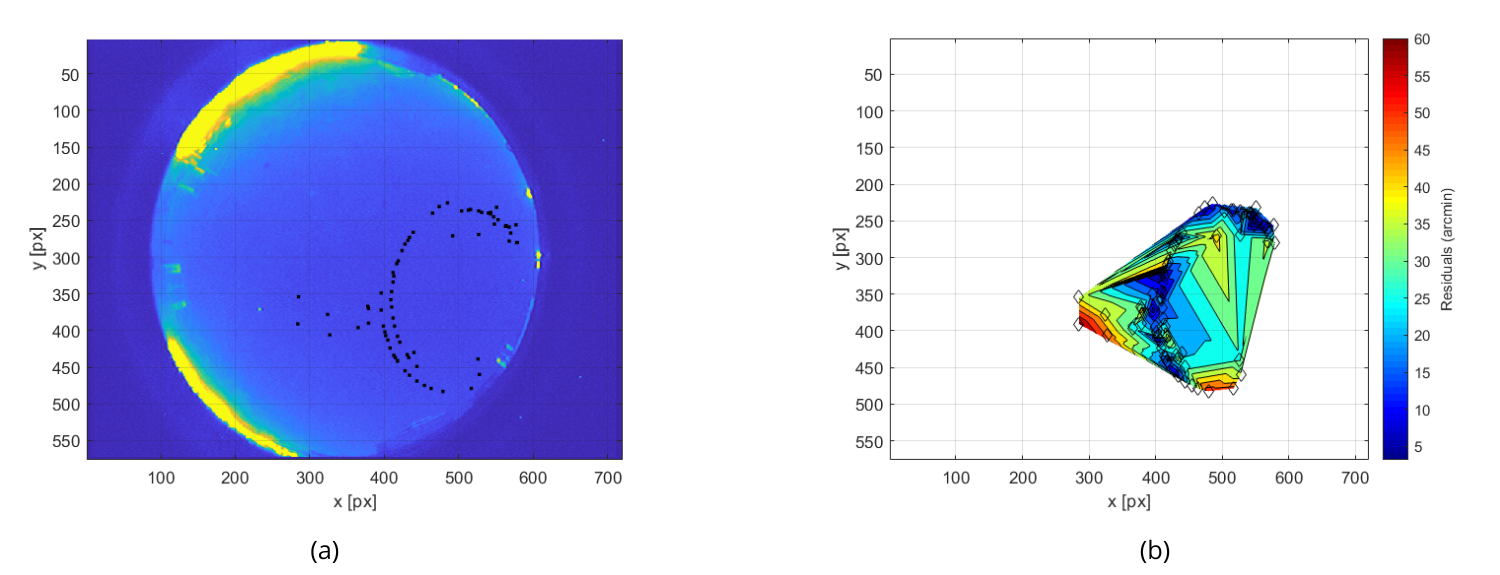}}
\caption{Datapoints used for astrometric calibration of Watec camera, indicated as black dots on false-color frame (a), and as magenta diamonds on filled contour of catalogue-vs-calculated error in arcminutes (b).}
\label{fig:dp_Watec}
\end{figure}

\begin{figure}[!ht]
\centering
{\includegraphics[width=\textwidth]{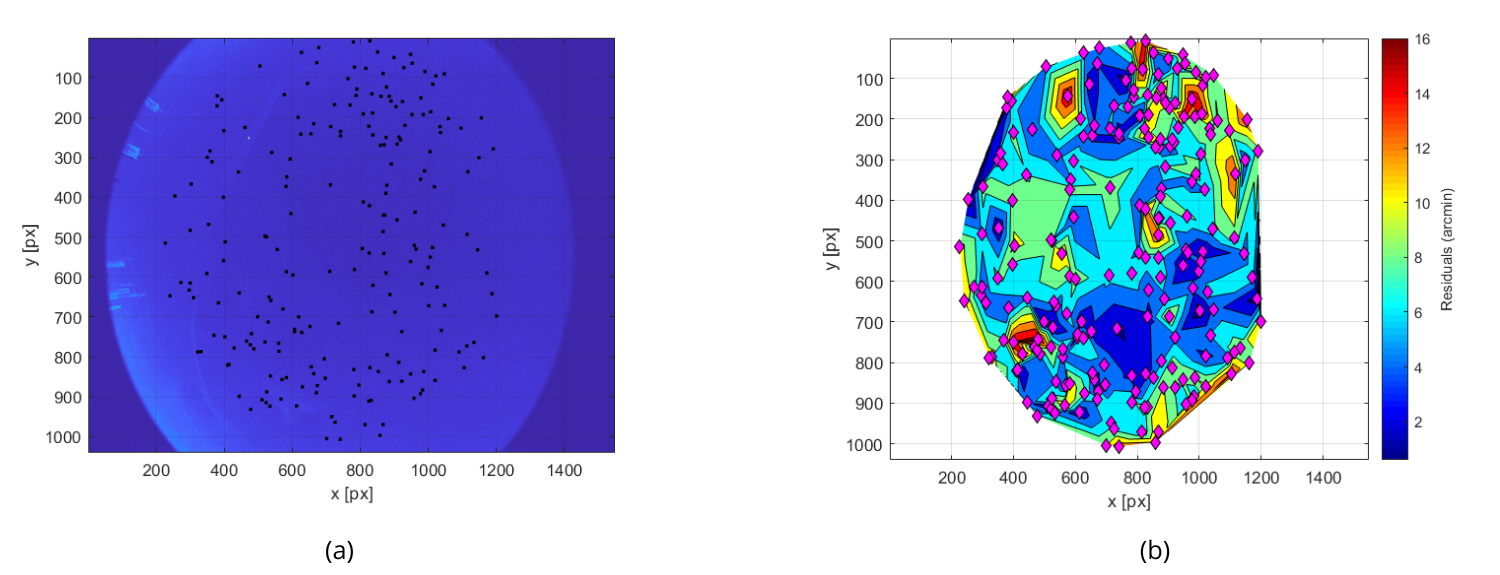}}
\caption{Data points used for astrometric calibration of ASI camera, indicated as black dots on false-color frame (a), and as magenta diamonds on filled contour of catalogue-vs-calculated error in arcminutes (b).}
\label{fig:dp_ASI}
\end{figure}

The quality metrics of the reduction are shown in Table \ref{table:res_astrom}, for both cameras. The parameters obtained using the subset are shown in Table \ref{table:astroparam}. The residuals of the astrometric reduction, calculated as: 

\begin{equation}
        \sigma_{res}=\sqrt{(z_i-z_{i,CAT})^2+((a_i-a_{i,CAT})\cdot\sin{z_{i,CAT}})^2}
        \label{eq:res_astrom}
\end{equation}

are shown in Figures \ref{fig:residuos_astrom} (a) and (b), for Watec and ASI, respectively. The median of the residuals is 22.3 arcmin for Watec, and 6.9 arcmin for ASI. The latter is consistent with the results reported in \cite{Vida2021} for GFO all-sky cameras using the model of \cite{Bannister2013} (pixel scale 1.9 arcmin/px, RMS residual 6.58 arcmin, see their Table A.3).
As expected, the performance of the ASI camera is a factor of three more accurate than the Watec, which is not only due to a factor of 4 higher resolution ($\sim$415 kpix vs. 1.61 Mpix) but also to the fact that significantly more datapoints are available when using the ASI camera, and with much lesser number of videos (less human working hours are needed to generate them). The datapoints being used are shown in Figures \ref{fig:dp_Watec} and \ref{fig:dp_ASI} for Watec and ASI, respectively. These figures show not only the higher number of datapoints in ASI with respect to Watec, but also the huge difference in image coverage for either system. The highest number of available datapoints, as well as their distribution in the image, for a lesser number of integrated frames, is of course a consequence of a higher sensor sensitivity in the case of ASI as compared to Watec, a fact that becomes clear in the photometric comparison that follows.  

\begin{table*}[ht!]
\centering
        \caption{Astrometric reduction qualilty metrics for the two evaluated cameras. Last column indicates average pixel resolution obtained from the reduction coefficient $p{_1}$.}
        \label{table:res_astrom}
        \begin{tabular}{|c|c|c|c|c|c|}
        \hline
        Camera & MAE & MAE, z & MAE, Az  & N$_{*}$ & Av. resol. \\
        &[arcmin]&[arcmin]&[arcmin]&&[arcmin/pix]\\
\hline
Watec & 15 & 15 & 15 & 74 & 14.8\\
ASI & 5 & 6 & 4 & 217 & 6.2\\
\hline
        \end{tabular}
\end{table*}

\begin{table*}[ht!]
\centering
        \caption{Astrometric reduction parameters for the two evaluated cameras.}
        \label{table:astroparam}
        \begin{tabular}{|c|c|c|c|c|c|c|c|c|}
        \hline
         & $x_0$ & $y_0$ & $p_0$ & $p_1$ & $p_2$ & $a_0$ & $E$ & $\epsilon$\\
         Camera  & [pix] & [pix] & [deg] & [arcmin$\cdot pix^{-1}$] & [arcmin$\cdot pix^{-2}$]& [deg] & [deg] & [deg]\\
\hline
Watec & 313.57  & 324.40  & 4.28  & 14.8  & 2.70$\cdot 10^{-4}$ & 102.95 & 222.22 & -7.51\\
ASI &  504.05 & 742.49  & 0.08 & 6.2  & 1.70$\cdot 10^{-5}$ & 89.12 & 223.47 & -1.20\\
\hline
        \end{tabular}
\end{table*}

\emph{Photometric comparison}. Using the previously found astrometric reduction parameters, we performed photometric reductions using our application as described in Section \ref{sec:photocalib}. The videos used for each camera were practically simultaneously acquired (2022/09/28 22:27:16, local time). The Watec video is 30s long (749 frames), while the ASI video is 10 s long (159 frames). We integrated the whole of each video streams, and searched for the 100 brightest stars present in our field. We narrowed down this list by considering a limiting altitude of 20$^{\circ}$ and a minimum S/N on the stars of 2 for both ASI and Watec. The minimum number of datapoints imposed on the iterative reduction process was 5 for Watec, and 20 for ASI. Aperture radii used were 2 and 10 pixels for r$_*$ and r$_{bgd}$, respectively, in the case of ASI, and 2 and 6, respectively, for Watec.

\begin{figure}[!ht]
\centering{\includegraphics[width=\textwidth]{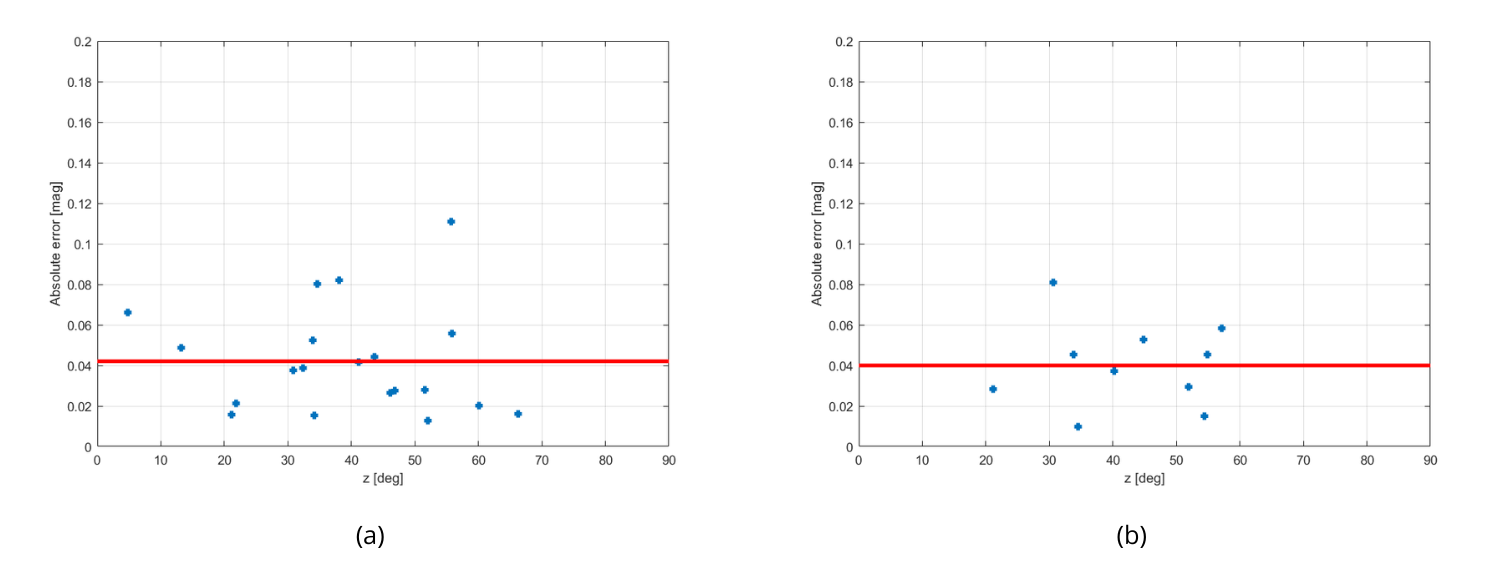}}
\caption{Residuals of the photometric reduction for Watec (a) and ASI (b) cameras. The red line indicates the mean of the residuals.}
\label{fig:residuos_fotom}
\end{figure}

Figure \ref{fig:residuos_fotom} shows the residuals of the photometric reduction for both cameras. The obtained reduction parameters are shown in Table \ref{table:fotom_results}, along with mean of the residuals. Also, estimates of limiting magnitudes (considering S/N of 5) obtained from linear fittings of catalogue magnitudes against the 10-based logarithm of the computed S/N are shown. For a single frame we obtain a limiting magnitude of +2.12, for ASI, and -1.58, for Watec.   

These results show that for shorter integration times, the ASI camera yields practically the same average residual as the Watec camera, with twice as much datapoints. Now, as stated before, our main goal is to process detected fireballs, which by definition are brighter than Venus ($m_V<-4$). Furthermore, in order to be able to perform photometry on these bright objects, which appear in videos of typically less than 5 seconds length, one can either use the same video for the photometric reduction, or generate a longer video right after the detection was triggered and the acquisition finished. The latter case generates a problem in automatic detection systems, since it more than duplicates the data volume in each station, which is critical when dealing with some tens of stations. This would be a problem even if using Watec cameras, that generates lighter video files than ASI. For this reason, using the same video, we performed an integration of 2 seconds for each camera (yielding 50 frames for Watec, and 30 for ASI), and applied the reduction parameters to compute the brightness of Jupiter ($m_V=$-2.935, according to the JPL Horizons ephemeris tool), which was present in the field. Besides the fact that Jupiter's brightness obtained with ASI was more accurate ($|\Delta m|=0.025$ mag.) than the one obtained with Watec ($|\Delta m|=0.625$ mag.), the reduction in the latter case could be performed with only 6 stars. 

This extrapolation to brighter objects is evaluated in more detail in next section, just for the ASI camera.

\begin{table*}[ht!]
\centering
        \caption{Photometric reduction qualilty metrics for the two evaluated cameras.}
        \label{table:fotom_results}
        \begin{tabular}{|c|c|c|c|c|c|c|c|c|}
        \hline
        Camera & MAE & m$_0$ & k & N$_*$  &  Lim. mag. (SNR>5)\\
        & [mag.] & [mag.] & [mag./AM] & & [mag.]\\
\hline
Watec & 0.04 & 6.09  & -0.261 & 10 & 2.79\\
\hline
ASI & 0.04 &  7.44 & -0.02& 20 & 2.76\\
\hline
        \end{tabular}
\end{table*}

\subsection{Validation of fireball photometry procedure}\label{sec:valid}
Once a photometric reduction has been performed, the post-processing application enables to perform flux measurements of bright, extended and (mostly) saturated objects, like fireballs but also stationary objects like planets and the Moon, and in fact any moving or stationary object present in the field.

The photometric comparison between the two cameras in the previous section was made using V magnitudes from the Hipparcos catalogue, and this was made so mainly because the stars used for Watec were $\alpha$-Cru, $\beta$-Cru, $\gamma$-Cru, $\alpha$-Cen and $\beta$-Cen, which are not available in the Gaia DR3 catalogue\cite{Gaia2022}. However, as it is shown in Figure \ref{fig:qe}, the spectral response of the CMOS sensor of the ASI camera is broader than the response of Hipparcos $V_T$ bandpass filter. For this reason, photometric reduction for ASI is more properly done using G magnitudes, for which we use a list of stars generated from the Gaia catalogue with m$_G\leq$3.5 and $\delta\leq$+50$^{\circ}$ \cite{Gaiab,Gaia2022}.

Nevertheless, it must be pointed out that the quality of photometric reductions performed using Gaia G magnitudes do not improve dramatically as compared to the ones obtained using $V_T$ magnitudes from Hipparcos, and in fact in some cases one can get lower MAE of calculated-vs-observed magnitudes using V than using G data. This is possibly due to the fact that what appears to dominate the accuracy of the photometric reduction is not what catalogue magnitudes are being used but possibly the intrinsic photonic noise of the stars that are being used for the reduction, which are not necessarily the same in one case and the other, since this depends on the iterative outlier removal process.

The application firstly computes the centroid of the extended source, and the user has to indicate its presence in the first frame only, the centroid in the remaining frames is automatically found by the program.

\begin{figure}[!ht]
\centering
{\includegraphics[width=\textwidth]{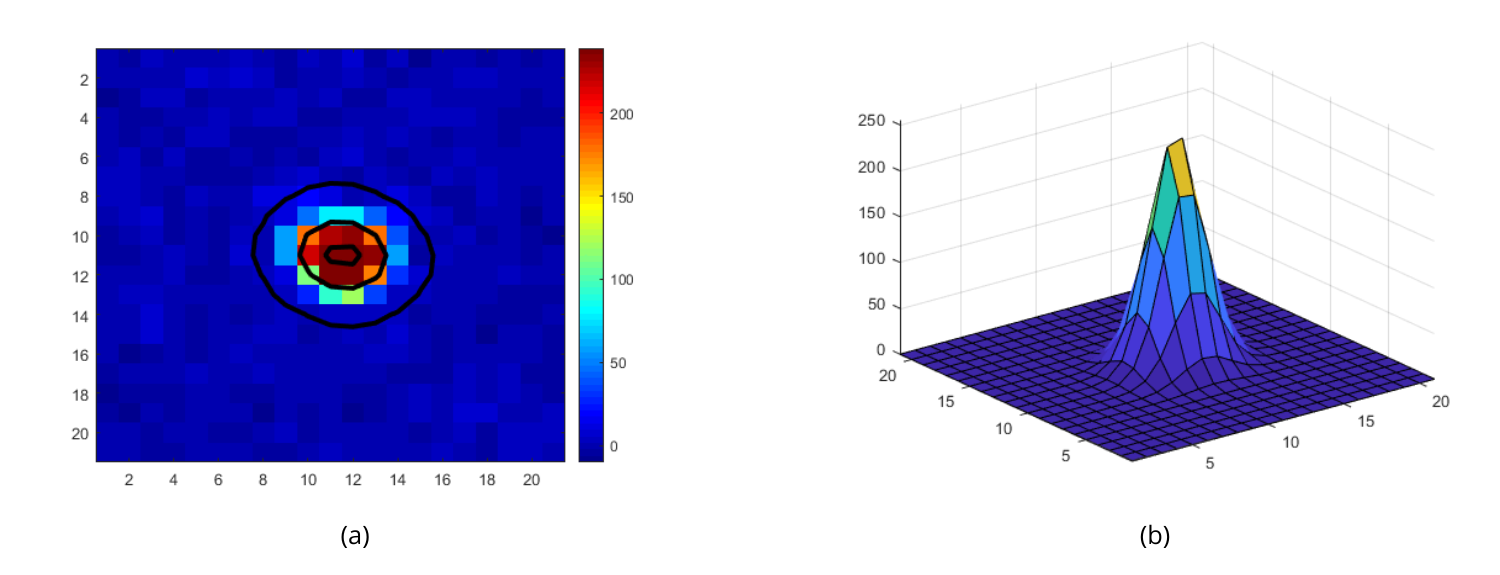}}
\caption{(a) Jupiter's sky-subtracted brightness and the fitted Gaussian's 3D plot (b). Black contour curves in (a) indicate 0.9, 0.5 and 0.05 of peak value of the fitted Gaussian.}
\label{fig:jupiter}
\end{figure}

Next, a two-dimensional Gaussian fit is made on the sky-subtracted brightness of the object in each frame, and the flux is spatially integrated. The sky background is computed as the mode of the unsaturated pixels present in a square centered on the object, of a user-defined size. Finally, this integrated, extrapolated flux is translated into an apparent magnitude using the previously found photometric reduction parameters. Also, the center (x$_G$,y$_G$) of the Gaussian is used to compute the astrometric position of the object in the different frames. Figure \ref{fig:jupiter} shows (a) the two-dimensional brightness of Jupiter in a single frame, and (b) the fitted Gaussian to the sky-subtracted brightness. Note that in Figure \ref{fig:jupiter} (a) several pixels are saturated.

\begin{figure}[!ht]
\centering
{\includegraphics[width=\textwidth]{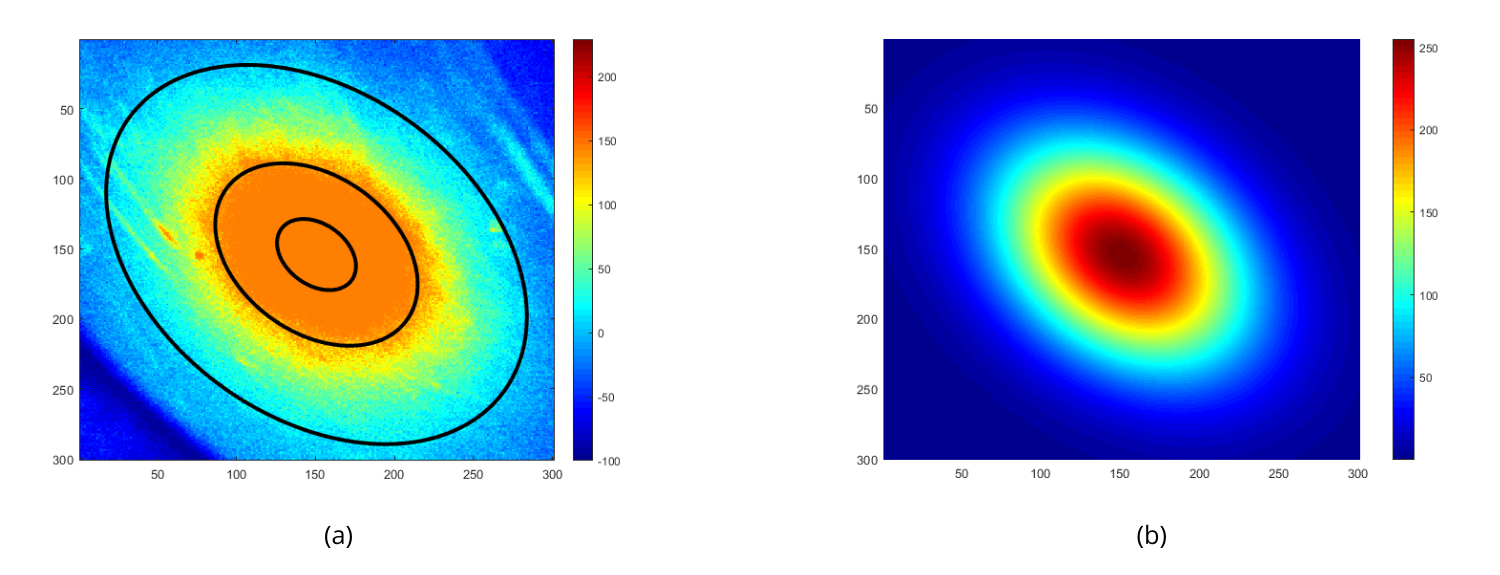}}
\caption{(a) Full Moon's sky-subtracted brightness and the corresponding fitted Gaussian's contour plot (b). Black contour curves in (a) indicate 0.9, 0.5 and 0.05 of peak value of the fitted Gaussian.}
\label{fig:luna}
\end{figure}

The results for the brightness estimate of Jupiter performed on the night of September 28th-29th, 2022, are summarized in Table \ref{table:jup_valid}. The errors reported in the third column of this table are with respect to Jupiter's V-magnitude computed with JPL Horizon Ephemeris Tool, and the relative MAE of these errors is $\sim$6\%. It does not appear to exist a systematic difference between the computed G magnitude and the corresponding V brightness of the planet. The fourth column shows the MAE of the residuals in the corresponding photometric reduction. In all cases, this reduction was performed with 20 stars.    

\begin{table*}[ht!]
\centering
        \caption{Results for the validation  of the photometric methodology applied to Jupiter, on the night of September 28th-29th, 2022. Mean brightness of Jupiter was -2.93, according to JPL Hporizons Ephemeris Tool. Third column indicates the difference between the computed magnitude, and the JPL V-magnitude. Fourth column indicates the MAE of calculated-vs-catalogue star magnitudes used in the photometric reduction.}
        \label{table:jup_valid}
        \begin{tabular}{|c|c|c|c|}
        \hline
        Local Time & z [deg.] & m-m$_{V,JPL}$ & MAE$_{cal}$\\
         \hline
         21:22:18 & 54.8 & -0.26 & 0.085\\
         22:27:16 & 44.3 & -0.04 & 0.186\\
         23:26:31 & 37.3 & +0.00 & 0.153\\
         00:13:48 & 34.4 & +0.10 & 0.156\\
         01:21:56 & 36.8 & -0.20 & 0.203\\
         02:52:03 & 48.4 & -0.18 & 0.077\\
         03:56:39 & 59.5 & -0.45 & 0.052\\
         \hline
         & MAE & 0.18 & 0.13\\
         \hline
        \end{tabular}
\end{table*}

The same procedure was applied to the full Moon of the night of November 7th-8th, 2022. The sky-subtracted brightness in a frame and the contour plot of the corresponding Gaussian fit are shown in Figures \ref{fig:luna} (a) and (b), respectively. As shown in Table \ref{table:luna_valid}, we obtain a relative MAE of $\sim$8\% with respect to the Moon's V magnitude. In this case, it does appear to exist a systematic difference between the computed G magnitude and the corresponding V brightness of the Moon, namely the computed G magnitude is higher (fainter) than the catalogue V magnitude. Since the sensor's bandpass is wider than a V filter (see Figure \ref{fig:qe}), this feature is not likely due to intrinsic radiation properties of the source and the detector's response to it. More likely, it is rather due to an over subtraction of sky background and the fact that since the source is extremely saturated the gaussian fit could be heavily biased, and this constitutes a challenge in such extended and saturated sources, such as the full Moon or a superbolide. 

It is not uncommon that panchromatic absolute magnitudes (magnitude at a distance of 100 km) are reported without an associated uncertainty. Exceptions are for instance FRIPON photometric data, which is reported with an uncertainty of half a magnitude \cite{Drolshagen_2020}.  Nevertheless, the relative error between computed G and catalogue V magnitudes shows that the procedure is suitable enough for obtaining light curves even for rare superbolides. In particular, it will enable us to determine whether its brightness is high enough to correspond to a massive enough object, to produce meteorites.     

\begin{table*}[ht!]
\centering
        \caption{Results for the validation  of the photometric methodology applied to the full Moon on the night of November 7th-8th, 2022. Mean brightness of the Moon was -12.63, according to JPL Horizons Ephemeris Tool. Third column indicates the difference between the computed magnitude, and the JPL V-magnitude. Fourth column indicates the MAE of calculated-vs-catalogue star magnitudes used in the photometric reduction.}
        \label{table:luna_valid}
        \begin{tabular}{|c|c|c|c|}
        \hline
        Local Time & z [deg.] & m-m$_{V,JPL}$ & MAE$_{cal}$\\
         \hline
         21:37:49 & 61.0 & +0.86 & 0.04\\
         22:51:23 & 53.4 & +1.39 & 0.06\\
         00:35:34 & 51.0 & +1.33 & 0.08\\
         01:33:03 & 54.7 & +1.15 & 0.16\\
         \hline
         & MAE & 1.18 & 0.09\\
         \hline
        \end{tabular}
\end{table*}

\subsection{First detection results}
\label{sec:firstdet}
\begin{figure}
    \centering
    \includegraphics[width=0.9\textwidth]{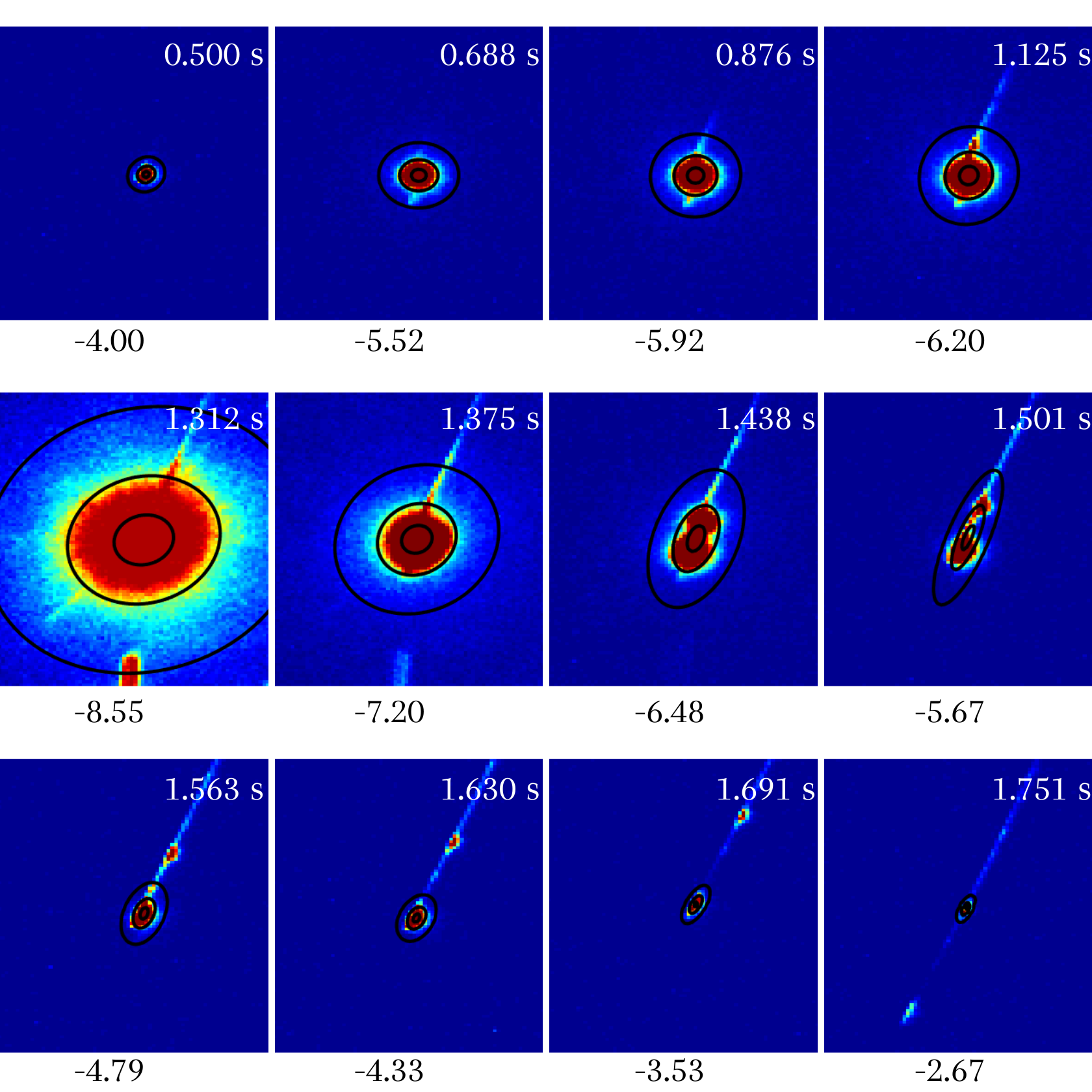}
    \caption{Fireball event of 29th October, 2022, UT 04:31:44, as observed from station 8. Apparent magnitudes are shown below each image, and times relative to first frame of the video stream are shown in upper right corner. Each box is 100x90 pixels.}
    \label{fig:bolido0}
\end{figure}

We present here the results of our postprocessing pipeline applied to a very bright fireball detected on 29th October, 2022, at UT04:31:44. We chose this event from a randomly selected sample of $\sim$1400 videos, spanning the $\sim$1 year operation of simultaneous operation of several stations of the network. These videos were  manually identified as fireball detection from a random sample of videos analyzed during our training workshops. We narrowed down the list to those events detected in at least 2 stations, and ended up with a list of about 20 videos. We selected this particular event due to its brightness and also because it presents clear evidence of fragmentation, as shown in the sequence of Figure \ref{fig:bolido0}.

The fireball was detected in five stations (1, 3, 4, 8 and 9), all located in the SE of Uruguay. At that point, stations 1-9 were operative. We did not include Station1 in the calculations that follow due to a low altitude of the fireball as observed from this station.

The astrometric reduction was done with stars present in event videos from within two nights from the fireball event. Each event video, typically $\sim$2s long, allows to identify $\sim$20-40 stars. Since we typically perform reductions on datasets of 150-200 datapoints, we usually need between 5 and 10 videos to produce this dataset, which are always available within $\sim$2 nights from the fireball event. For the photometric reduction, stars present in the fireball event are used.

We make use of the method of planes (MOP) described in \cite{Ceplecha1987}, for the multi-station case, which implies finding a weighted average trajectory and projecting observations -as seen from each station- on to it. The weight is the square of the sine of the angle between the planes considered for each triangulation, multiplied by the astrometric precision or the length of the observation, as detailed in the aforementioned work.

For each station $S$, based on these projected datapoints, we compute their geodetic coordinates, height $h$ above mean sea-level and absolute panchromatic magnitude $M$. Next, we sort all datapoints in ascending time order, and compute the distance $l$ traveled along the trajectory and average velocity $v$. 

We clean each time series by considering that $h(t)$ is a monotonically decreasing function over time, and $l(t)$ is a monotonically increasing function over time. 

\begin{figure}[!ht]
\centering{\includegraphics[width=\textwidth]{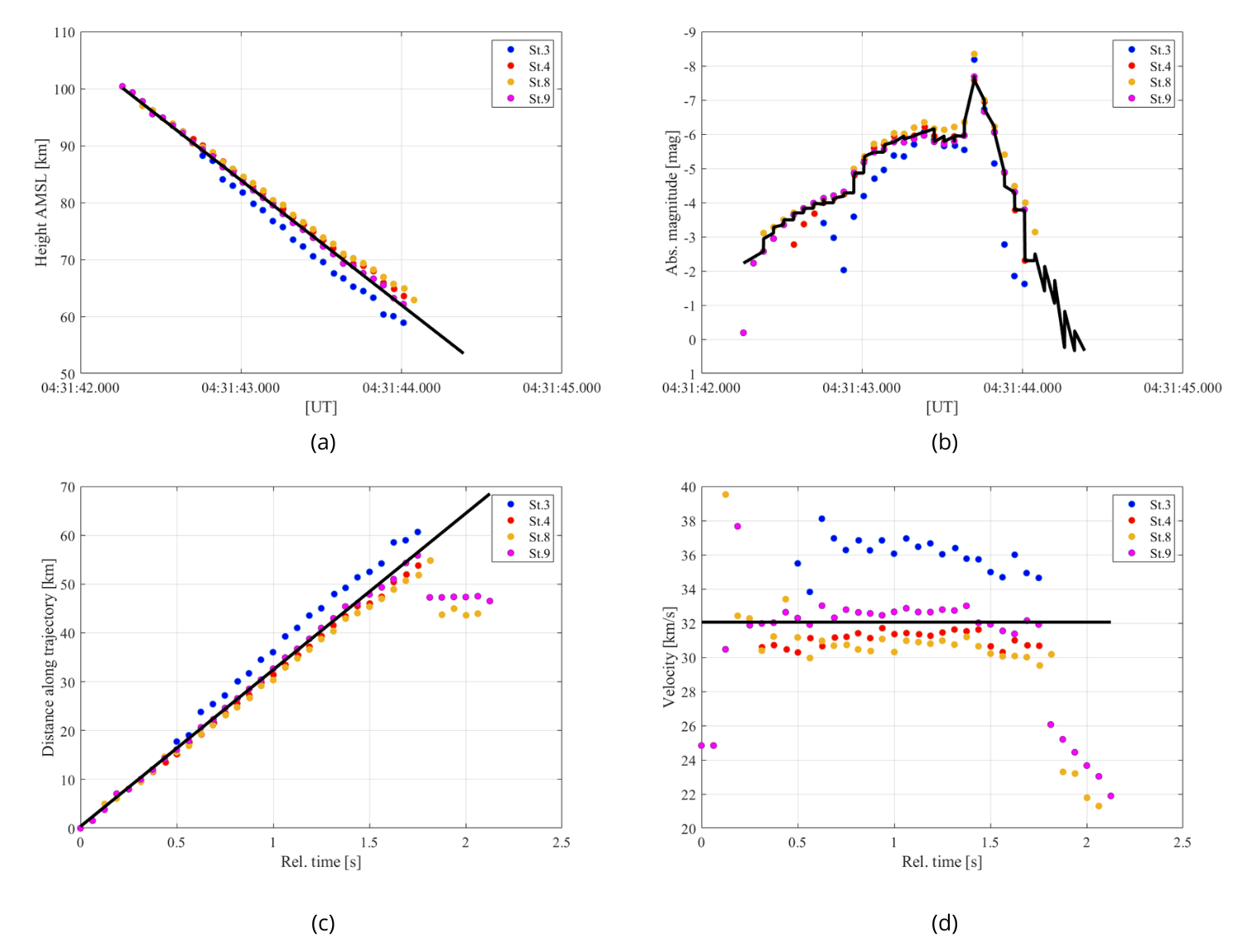}}
\caption{Time series of the 29/10/2022 fireball (solid black lines, see text for further details); a) Height above mean-sea-level. b) Absolute magnitude. c) Distance traveled along trajectory. (d) Average velocity. Scatter plots correspond to projection of observations on the weighted average trajectory as seen from stations 3 (blue), 4 (red), 8 (orange) and 9 (magenta).}
\label{fig:bolido1}
\end{figure}

$h(t)$ is shown in Figure \ref{fig:bolido1} (a). The solid black line is a linear fit on the datapoints. Figure \ref{fig:bolido1} (b) shows $M(t)$, and the solid black line in this case is the result of a moving median filter, with a 100 ms window. Figure \ref{fig:bolido1} (c) shows $l(t)$, and the solid black curve is a linear fit on the datapoints. One could include an exponential term (see \cite{Whipple_1957}) but for this short flight it does not improve the goodness of the fit. Finally, Figure \ref{fig:bolido1} (d) shows average velocities and in this case the horizontal black line indicates the mean velocity obtained from the linear fit on $l(t)$.

The outliers clearly seen in Figures \ref{fig:bolido1} (c) and (d) are due to the fact that, as seen in Figure \ref{fig:bolido0}, the centroid of the fireball during the fragmentation event (1.438s-1.501s) is located somewhere between the position of the two main fragments, then (1.563s) it "follows" the faster fragment and finally (1.751s) "goes back" to the slower fragment.

The fireball was detected at a height of 100.1 km$\pm$0.2km, and it was observed down to 53.5 km$\pm$0.1km. The peak absolute magnitude was -7.682$\pm$0.003, mean velocity 32.1 km/s$\pm$0.1km/s and longest observed duration of the flight 2.13 s. The uncertainties associated to each result are obtained from clone simulations based on our astrometric error in $z$ and $a$ (see Table \ref{table:res_astrom}).

We do not observe any evident trend of deceleration, which is expected since the deceleration at a given time instant is roughly given by (\cite{Halliday_1996}):
\begin{equation}
    \dot{v}=-\frac{1}{2}\frac{A}{m_d}\cdot \rho_a \cdot v^2
\end{equation}
where $A$ and $m_d$ are the dynamic cross-section and mass, respectively, $\rho_a$ the atmospheric density and $v$ the instantaneous velocity. Assuming, as in \cite{Halliday_1996}, a "brick"-shaped meteoroid with sides 2L, 3L and 5L (an assumption based on the recovered fragments of the Innisfree fall), the area-to-mass ratio $A/m_d$ becomes proportional to 1/L. At the initial height and inserting the mean velocity, and assuming some typical values for the remaining parameters (meteoroid density 3.5 g/cm$^3$, isothermal atmosphere with planetary atmospheric height scale $h_0=$7.16$\cdot$10$^5$cm) the deceleration is $\sim$10$^{-2}$/L km/s$^2$, where L is in cm. So for any larger than cm-sized meteoroids, at this height there is a negligible deceleration. In such cases, the mean velocity is a good estimate of the initial velocity of the meteoroid. 

\begin{figure}
    \centering    \includegraphics[width=0.9\textwidth]{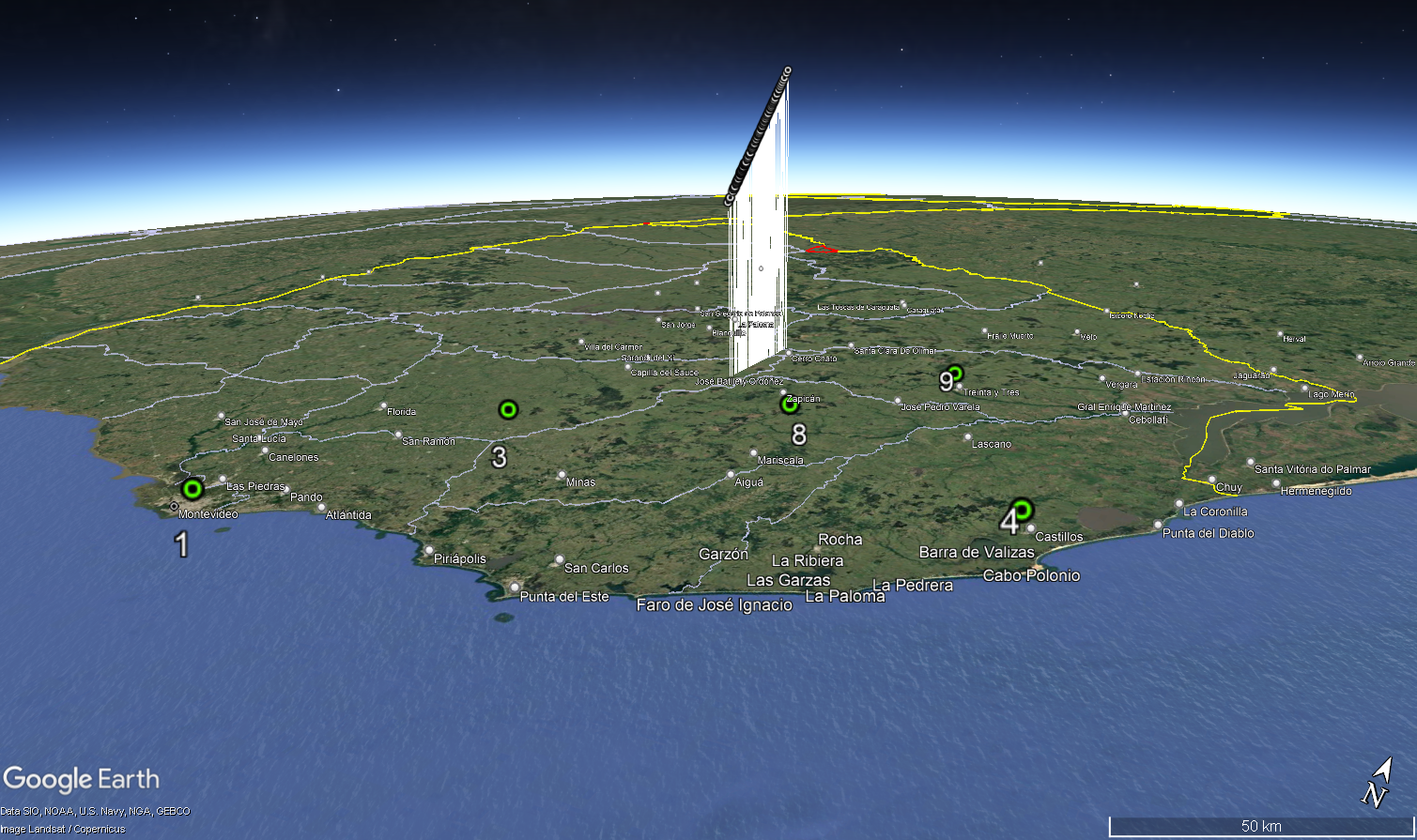}
    \caption{3D reconstruction of the fireball's trajectory obtained through the MOP of \cite{Ceplecha1987}, as described in the text (source of the map is Google Earth Pro$\copyright$ v7.3.6.10201, based on Landsat/Copernicus imagery of 12/31/2020, eye alt. 138.06km. SIO, NOAA, U.S. Navy, NGA, GEBCO).}
    \label{fig:bolido3}
\end{figure}

Geodetic coordinates of each data point's projection on the ground is easily found, and enables to reconstruct the meteoroids 3D trajectory, as shown in Figure \ref{fig:bolido3}.

Following \cite{Ceplecha1987}, we found the average geocentric radiant and velocity, which are listed in Table \ref{table:events}.

\begin{table*}[ht!]
\centering
        \caption{Average geocentric radiant coordinates and velocity.}
        \label{table:events}
        \begin{tabular}{|c|c|c|c|c|}
        \hline
        Date & UT & $\alpha_{G,J2000}$ & $\delta_{G,J2000}$ & v$_G$\\
         & & [$^\circ$] & [$^{\circ}$] &[km/s]\\
         \hline           
         2022-10-29 & 04:31:44 & 57$\pm$1 & +14.7$\pm$0.1 & 29.7$\pm$0.2\\
         \hline
        \end{tabular}
        
\end{table*}

\cite{Jenniskens_2016} report nineteen components of the Encke complex, which consists of a Southern and a Northern branch. Our particular event matches the Southern Taurids, and we could possibly associate it to the STS (IAU 628) component ($\alpha_G$=53.8$^\circ$, $\delta$=+14.4$^{\circ}$, v$_G$=28.2 km/s).

Initially associated to the comet 2P/Encke, the Taurid complex is now believed to consist of a family of Near-Earth Objects together with meteoroid streams (see e.g. \cite{Jenniskens_2016}$^,$\cite{Pena-Asensio_2021} and references therein). Therefore, the established association does not rule out an asteroidal nature of the meteoroid. Nevertheless, the relatively high end altitude does rule out the possibility of being a meteorite-dropper (see e.g. \cite{MorenoIbanez_2015}).

\subsection{Future perspectives}
Although we had finished the installation of the 20 stations in February 2023, we are continually maintaining and upgrading the equipment, with the support of the local teachers and students. Improvements in the different software tools (control application, status monitoring and post-processing) are being done and uploaded to each station.

The data handling is challenging, as has already been noted with the current state of the network. This imposes the need for an efficient automatic video classifier (as in e.g. \cite{Anghel_2022}$^,$ \cite{Gural_2020}), which has been developed and tested on a testing sample of Watec videos, but not run on the entire dataset, which comprises videos acquired both with Watec and ASI cameras. Once the testing and classification of the whole dataset to date has been carried out, results of both the classifier's performance and the statistics of the data will be presented in a future publication.   

The mapping function used in the astrometric reduction process leads to an increasing error with decreasing altitude, which indicates that is not suited for the lens being used. Any other available model showing better performance at low altitudes could be implemented (e.g. the GMN $r^7$model of \cite{Vida2021}). 

Another future line of research that a classified dataset will allow us to explore is to look for detections of the brightest events in the Geostationary Lightning Mapper (GLM) onboard the GOES-16 satellite, as reported in \cite{jenniskens17}, which will add additional data for trajectory and orbit reconstruction.

Finally, it is to be undertaken the task of implementing the dark flight stage in our data processing pipeline. The strato- and tropospheric wind data is a challenge, since the closest meteorological stations having open and available data (e.g. at the website of the Department of Atmospheric Science, University of Wyoming\cite{Wyoming}) are Santa Rosa in Argentina (87623 SAZR, $\phi$=-36.56,$\lambda$=-64.26, ca 750km SW from Montevideo) and Santa Mar\'ia in Brasil (83937 SBSM, $\phi$=-29.72,$\lambda$=-53.70, ca 600km NE from Montevideo). The data from these stations may be useful for events close to the Argentinian and Brazilian borders, respectively. In addition, atmospheric data from the Global Forecast System of NOAA (GFS\cite{GFS}) may be used in any case, as described in \cite{Moilanen2021}. Also, the local national weather service INUMET has a network of $\sim$40 automated meteorological stations with available wind data at ground level, which may complement the former data on higher altitude winds\cite{INUMET}. It is to be investigated whether other data sources could be available on request.

\section{Conclusions}\label{sec:concl}
We recently finished (February 2023) the deployment of a nationwide network of 20 all-sky stations for automatic detection of fireballs in Uruguay, South America, as a contribution to the global deployment of automated fireball networks and to the increase of the number of recoverable meteorite falls. 

As a citizen science project it involves secondary school students and teachers in the operation of the stations and the analysis of the data.

The hardware is of local design, and software has also been developed locally for the automatic detection and post-processing of fireballs, and are freely available in \emph{github}. A system was also implemented to report the status of all stations in the network in real time, and the information is displayed on a public website.

Our post-processing pipeline permits to find the trajectory in horizontal coordinates of detected events, through an astrometric reduction with accuracies in good agreement with other works (e.g. \cite{Vida2021}), and process multistation observations through the MOP triangulation method of \cite{Ceplecha1987}. 

Since initially the network was based on the widely used Watec 902 H2 ultimate camera, and it was later replaced with a higher resolution camera, ZWO ASI 178MM, a comparison between the astrometric and photometric performances of both systems has been carried out. 

The comparison shows, as expected for a higher resolution sensor, a better astrometric performance for the ASI, achieving a mean absolute error of  $\sim$5', while for Watec we obtain $\sim$15'. As important are the photometric results of this comparison, in particular in the fact that due to the higher sensitivity of the ASI's sensor, a fireball video (usually 1-3 seconds long) can itself be used to perform the photometric reduction with the stars present in the field, something which is barely possible with a video acquired with Watec, if possible at all. In the latter case, the acquisition software should automatically generate, right after the detection was triggered and the event generated, a longer video for photometric reduction, something which poses a challenge in terms of data handling and storage, a critical issue when dealing with a network of tens of stations. 

Also, we have implemented a methodology for finding the integrated flux of a fireball and estimating its magnitude, by extrapolating a photometric reduction performed with Gaia DR3 G-photometric star magnitudes. The method has been validated with bright, extended objects such as Jupiter and full Moon, and the obtained brightnesses have relative errors of 6 and 8\%, respectively.

With the network fully operational since February 2023, we are now working on the automatic classifier that will produce a database of registered meteors,  and also on the final stages of our data processing pipeline.

\subsection*{Disclosures}
The authors declare that there are no financial interests, commercial affiliations, or other potential conflicts of interest that could have influenced the objectivity of this research or the writing of this paper.

\subsection*{Data availability}
The data that support the findings of this article is not public yet since we are currently working on data storing issues, but can be requested from the author at manuel.caldas@fcien.edu.uy. 

\subsection*{Acknowledgments}

The deployment of the network would have been impossible without the support of the Direcci\'on General de Enseñanza Secundaria and Direcci\'on General de UTU (Uruguay), that allowed us to install the stations in 19 different High-Schools. In particular we thank: Reina Reyes, Mar\'ia Acuña (Inspecci\'on de Astronom\'ia, Direcci\'on General de Enseñanza Secundaria), Ra\'ul Salvo (Observatorio Astron\'omico de Montevideo), Carlos Cladera, Angel Ceballos (Liceo 1 San Carlos, Maldonado), Heber Pombo, Mateo Muñoz (Liceo Casup\'a, Florida), Carolina Nicodela (Liceo Castillos, Rocha), Cinthia Davila, Ver\'onica Orosco, Marina Rohrer, Julia Cardozo (Liceo Rosario, Colonia), Alejandra Banchero, Neville Charbonnier, Freddy Planchon (Liceo Dolores, Soriano), Juan Bordon, Fabian Regalado (Liceo 1 Trinidad, Flores), Jose Rossy, Martin Zuarez (Escuela Agraria Piraraj\'a UTU, Lavalleja), Sandra Conde, Santiago Silva (Liceo 1 Treinta y Tres, Treinta y Tres), Gabriela Talice (Liceo San Gregorio de Polanco, Tacuaremb\'o), Christian L\'opez, Ana Laura Cabrera (Liceo 1 Melo, Cerro Largo), Henry Olivera (Liceo Rio Branco, Cerro Largo), Mart\'in Lacuesta (Liceo 1 Paysand\'u, Paysand\'u), Nelson Causa, Camila Negri (Liceo Guich\'on, Paysand\'u), Diego Silveira (Liceo 1 Tacuaremb\'o, Tacuaremb\'o), N\'estor Bustamante (Liceo Vichadero, Liceo 5 Rivera, Rivera), H\'ector Rodr\'iguez, Ra\'ul Cavallo (Liceo 1 Salto, Salto), Nelson Cooper (Liceo 2 Artigas, Artigas), Jorge Larrosa (Liceo 2 Bella Uni\'on). Also, we had the support of the different institutions' Directors, administration and technical staff.

We acknowledge the work of more than 30 high-school students and teachers that attended the training workshops, and helped us in the analysis of the data (see the full list at: \url{http://bolidos.astronomia.edu.uy/taller-bocosur/}).

Also, we thank the Comisi\'on Sectorial de Investigaci\'on Cient\'ifica (Udelar, Uruguay), Programa de Desarrollo de las Ciencias B\'asicas (PEDECIBA, Uruguay), and the Embassy of the United States of America for their financial support.

We thank Ignacio Ramirez (Facultad de Ingenier\'ia, Udelar, Uruguay) for an initial version of the \emph{Sentinela} python script for monitoring the status of the stations.

We thank Peter Gural for his suggestions on meteor photometry. 

This work has made use of data from the European Space Agency (ESA) mission
{\it Gaia} (\url{https://www.cosmos.esa.int/gaia}), processed by the {\it Gaia}
Data Processing and Analysis Consortium (DPAC,
\url{https://www.cosmos.esa.int/web/gaia/dpac/consortium}). Funding for the DPAC
has been provided by national institutions, in particular the institutions
participating in the {\it Gaia} Multilateral Agreement.

%%%%% References %%%%%

\bibliography{report}   % bibliography data in report.bib
\bibliographystyle{spiejour}   % makes bibtex use spiejour.bst

%%%%% Biographies of authors %%%%%

\vspace{2ex}\noindent\textbf{Manuel Caldas} is an Assistant Professor at the University of the Republic of Uruguay. He received his BS degree in Electrical Engineering from Chalmers University of Technology in 2007 and MS in astronomy from the University of the Republic 2012, and presented his PhD thesis in Physics Engineering from the University of the Republic in november, 2024. His current research interests include meteor science and radio astronomy.

\vspace{2ex}\noindent\textbf{Luc\'ia Velasco, Valeria Abraham and Alvaro Guaimare} are Master's students at the Astronomy Department of the Faculty of Sciences of the University of the Republic of Uruguay.

\vspace{2ex}\noindent\textbf{Lucas Barrios and Mat\'ias Hern\'andez} are undergraduate students at the Astronomy Department of the Faculty of Sciences of the University of the Republic of Uruguay.

\vspace{2ex}\noindent\textbf{Gonzalo Tancredi} is a Professor at the Astronomy Department of the Faculty of Sciences of the University of the Republic of Uruguay. His current research areas are impacts in granular media, planetary defense and meteor science.

\listoffigures
\listoftables

\end{spacing}
\end{document}